%% file: main.tex
\documentclass{article}

 \usepackage[preprint]{neurips_2026}

\usepackage[utf8]{inputenc} 
\usepackage[T1]{fontenc}    
\usepackage{hyperref}       
\usepackage{url}            
\usepackage{booktabs}       
\usepackage{amsfonts}       
\usepackage{nicefrac}       
\usepackage{microtype}      
\usepackage{xcolor}         
\usepackage{amsmath,amssymb}
\usepackage{algorithm}
\usepackage{algorithmic}
\usepackage{comment}
\usepackage{graphicx}
\usepackage{color}
\input{math_commands}

\title{Emergence of Social Reality of Emotion through a Social Allostasis Model with Dynamic Interpretants}

\author{%
  Kentaro Nomura \\
  Graduate School of Engineering Science\\
  The University of Osaka\\
  \texttt{nomura.kentaro.7ks@ecs.osaka-u.ac.jp} \\
  \And
  Yushi Tsubamoto \\
  Graduate School of Medicine\\
  The University of Osaka\\
  \texttt{u935741h@alumni.osaka-u.ac.jp}\\
  \And
  Takato Horii \\
  Graduate School of Engineering Science\\
  The University of Osaka\\
  \texttt{takato@sys.es.osaka-u.ac.jp} \\
}

\begin{document}

\maketitle

\begin{abstract}
The theory of constructed emotion defines social reality as the community-level consensus on emotion concepts assigned to interoceptive sensations arising from bodily allostasis and social interaction.
In this study, we simulate this emergence process using a computational model that integrates symbol emergence with degrees of freedom in symbol interpretation and active inference.
Two agents receive interoceptive signals, exchange inferred symbols, and simultaneously adapt their bodily control goals and symbol interpretations to each other.
Experimental results show that the interoceptive prior preferences and symbol probability distributions of the two agents converge, confirming the emergence of social reality grounded in social consensus.
\end{abstract}

\section{Introduction}
\label{sec:introduction}

The theory of constructed emotion explains emotion as a phenomenon that emerges from the predictive processing of bodily states and communicative interaction with others~\cite{Barrett2017-cv,Barrett2025-vx}.
A central function of the brain in this framework is body budgeting: the brain continually predicts the body's future metabolic demands on the basis of interoceptive sensations and proactively allocates resources to meet those demands, a process known as allostasis~\cite{barrett2017emotions}.
Affective experience arises in the course of this allostatic regulation as a low-dimensional summary of the body's current metabolic state~\cite{Barrett2017-cv}.
These affective experiences are then categorized into discrete emotional concepts through a predictive process in which the brain draws on prior experience and contextual information, including socially shared concepts such as language and symbols~\cite{Barrett2026-kq}.
Emotions are thus understood not as biologically fixed categories but as conceptual categories that are dynamically constructed through the interplay of bodily regulation and social interaction, and whose functions are conferred by collective agreement.
The community-level consensus on emotion concepts assigned to interoceptive sensations is termed the \emph{social reality} of emotion~\cite{Barrett2012-ik,Searle2010-cr}.
This paper proposes a computational model in which two agents mutually control their bodily states while communicating through symbols, and demonstrates that the social reality of emotion---a state in which both agents share aligned bodily control goals and co-constructed emotion concepts---emerges through this interaction.

The notion of social reality in the theory of constructed emotion draws on a broader philosophical tradition concerning the social constitution of reality.
Searle argues that institutional facts---facts that exist only by virtue of human agreement, such as money or marriage---are created through collective intentionality and constitutive rules of the form ``X counts as Y in context C''~\cite{Searle1995-jh}.
Berger and Luckmann offer a complementary account in which social reality is produced through a continuous cycle of externalization, objectivation, and internalization: individuals create shared concepts through interaction, these concepts become institutionalized and take on an objective quality, and subsequent members of the community internalize them as given~\cite{Berger1966-nm}.
Epstein further refines the analysis by distinguishing two relations that underlie social facts: \emph{grounding}, which determines what makes a particular social fact obtain, and \emph{anchoring}, which determines how a social category is set up in the first place~\cite{Epstein2015-yv}.
Applying this distinction to emotion, the grounding of an emotion instance lies in the individual's interoceptive state and its conceptual categorization, whereas the anchoring of an emotion category is provided by the community-level consensus that establishes the category as a shared concept.
In this view, the generation of social reality within a community is a co-constructive process in which emotion concepts are both constituted by and constitutive of the social interactions among community members.

Cultural and interpersonal contexts play a central role in this co-constructive process.
Research on cultural differences in emotions has shown that the most prevalent emotional phenomena in a culture are those that fit and reinforce that culture's models of self and relationship~\cite{Mesquita2003-he}.
Furthermore, emotions are co-constructed through interpersonal interactions and relationships embedded in broader cultural contexts, so that emotional experience is not merely an individual phenomenon but a socially negotiated process~\cite{Boiger2012-vp}.
These findings suggest that a comprehensive account of emotion must address both the bodily processes of interoceptive regulation and the social processes through which emotion concepts are shared and negotiated.

When emotions are viewed as symbols, their co-construction can be characterized as a bottom-up symbol emergence process driven by bodily and social interaction.
The collective predictive coding hypothesis explains symbol emergence as the formation of external representations through a group's adaptation to the environment via decentralized Bayesian inference, and the Metropolis--Hastings Naming Game (MHNG) has been proposed as a representative implementation~\cite{Taniguchi2024-ym,Taniguchi2023-lj}.
Computational models of emotion concept formation have demonstrated that emotion categories can be formed from multimodal sensory information including interoception through probabilistic generative models~\cite{Tsurumaki2025-vu}, and that agents can develop shared signs from multimodal observations through MHNG~\cite{Hoang2023-gc}.
On the bodily side, the Embodied Predictive Interoception Coding (EPIC) model formalizes allostasis as the control of bodily states through active inference of interoceptive sensations~\cite{Barrett2016-fz}.
However, each of these lines of research addresses either the social sharing of emotion concepts or the bodily regulation of interoceptive states, and a model that considers both aspects in an integrated manner has not been explored.

This paper proposes a computational model of the social reality of emotion that integrates two processes: the regulation of bodily states at the individual level and the social sharing of emotion concepts through symbolic communication.
In this model, two agents each receive interoceptive sensations and exchange symbols inferred from their internal states, while simultaneously adapting their bodily control goals and symbol interpretations to each other.
Through simulation experiments, we demonstrate that the agents' prior preferences over interoceptive sensations converge and that symbols acquire shared meanings representing bodily control actions, thereby confirming the emergence of the social reality of emotion from the interplay of bodily regulation and social interaction.

\section{Related Work}
\label{sec:related}

\subsection{Allostasis and Interoceptive Prediction in Emotion Construction}
\label{sec:related:allostasis}

A central mechanism in the theory of constructed emotion is allostasis, the anticipatory regulation of the body's internal milieu~\cite{Barrett2017-cv}.
Under this account, the brain maintains internal models that generate predictions about forthcoming bodily demands and regulates internal states accordingly; affective experience emerges in the course of this predictive regulation.
When these affective experiences are categorized through socially shared concepts, they become emotions.
Barrett et al. formalize this process as active inference of interoceptive sensations in the Embodied Predictive Interoception Coding (EPIC) model~\cite{Barrett2016-fz}.
The EPIC model provides a mathematical account of how the brain controls bodily states by minimizing prediction error with respect to interoceptive signals, thereby offering a computational basis for the bodily dimension of emotion construction.
However, the EPIC model focuses on the individual level and does not address the social processes through which emotion concepts are negotiated and shared within a community.

\subsection{Symbol Emergence and the Collective Predictive Coding Hypothesis}

Symbol emergence provides a framework for understanding how shared representations arise through bodily and social interaction.
Taniguchi proposes the collective predictive coding hypothesis, which characterizes symbol emergence as the formation of external representations through a group's adaptation to its environment~\cite{Taniguchi2024-ym}.
In this hypothesis, symbols are inferred via decentralized Bayesian inference.
The Metropolis--Hastings Naming Game (MHNG) has been proposed as a representative computational implementation of this process~\cite{Taniguchi2023-lj}.

\subsection{Computational Models of Emotion Concept Formation}

Computational approaches to the formation of emotion concepts from multimodal information are also relevant to the present study.
Tsurumaki et al. use multilayered multimodal latent Dirichlet allocation, a probabilistic generative model, to form emotion concepts from vision, physiology, and word information~\cite{Tsurumaki2025-vu}.
That model demonstrates that emotion categories consistent with human subjectivity can be formed through the integration of interoceptive and exteroceptive information.
In a related direction, Hoang et al. study emergent communication based on multimodal deep generative models and MHNG, showing that agents can develop shared signs from multimodal observations through decentralized Bayesian inference~\cite{Hoang2023-gc}.

Each of these prior studies, however, addresses only one aspect of the problem: the bodily dimension of interoceptive regulation (Section~\ref{sec:related:allostasis}), the formation of emotion concepts from multimodal information, or the social sharing of categories through symbolic interaction.
A model that jointly accounts for the bottom-up bodily process and the top-down social process remains unexplored.
The present study addresses this gap by proposing a model that integrates active inference for bodily state control with symbolic communication for sharing emotion concepts.

\section{Proposed Model}
\label{sec:model}

\begin{figure}[t]
    \centering
    \hspace{-11mm}
    \includegraphics[width=0.9\linewidth]{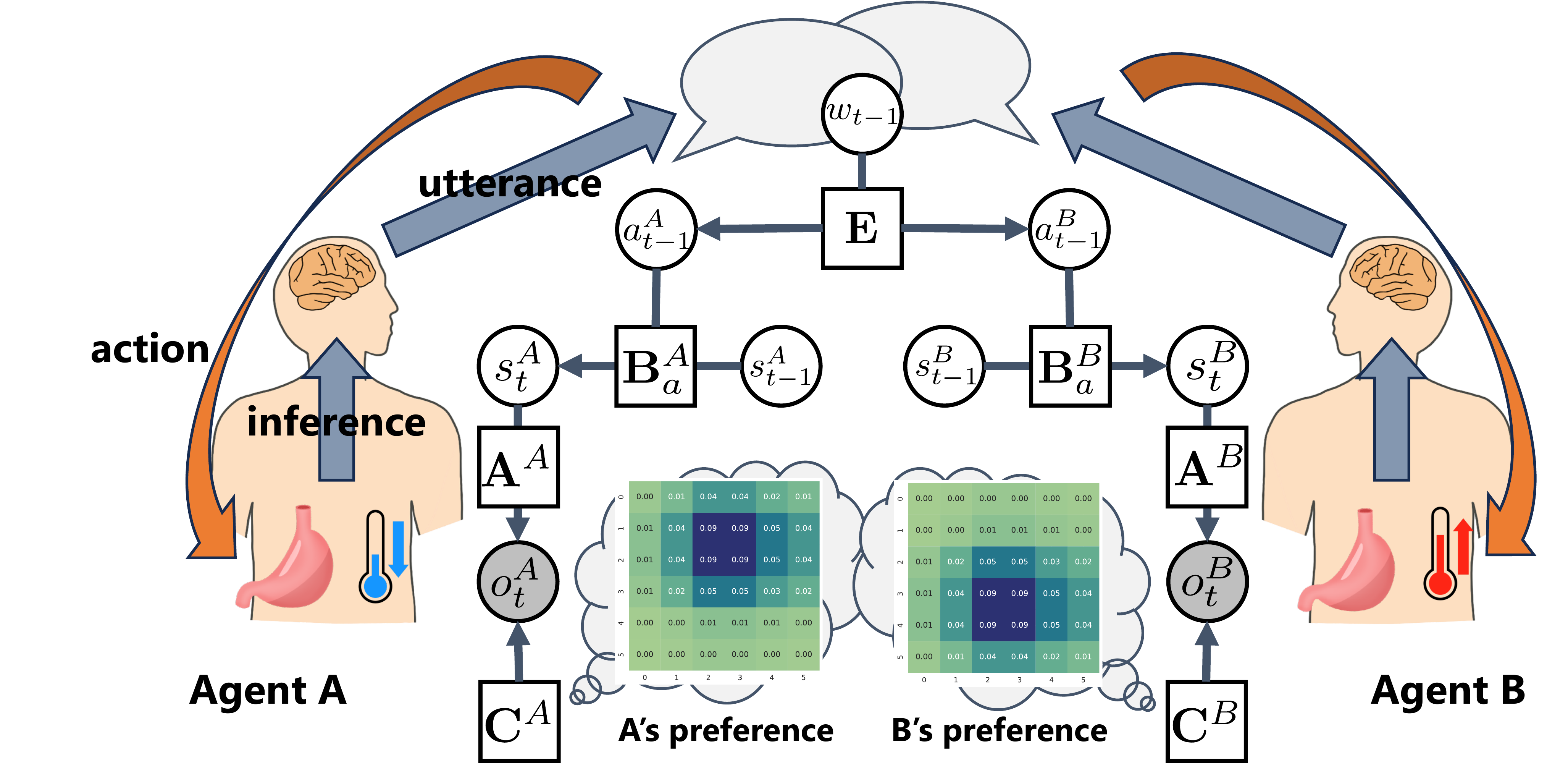}
    \caption{Overall architecture of the proposed model.
    Each agent maintains a POMDP-based generative model with likelihood $\mathbf{A}^*$, state transition $\mathbf{B}^*_a$, and prior preference $\mathbf{C}^*$.
    Agents exchange symbols through the shared symbol interpretation parameter $\mathbf{E}$ via the Metropolis--Hastings Naming Game and select actions to control their bodily states through active inference.}
    \label{fig:model_overview}
\end{figure}

The proposed model simulates the process in which social reality emerges as two agents jointly attend to each other's interoceptive sensations, exchange symbols inferred through active inference, select actions to control their bodily states, and simultaneously learn the dynamics of those states and the interpretation of symbols.
Figure~\ref{fig:model_overview} illustrates the overall architecture of the proposed model.
This section first describes the generative model of each agent and the symbol interpretation model shared by the community, then explains symbol inference and communication based on active inference and collective predictive coding, and finally presents the parameter update rules.

\subsection{Discrete Model of Interoception}
\label{sec:model:discrete}

Each agent possesses its own internal model for generating and inferring interoceptive sensations, formulated as a partially observable Markov decision process (POMDP) with discrete states and discrete observations.
In addition, both agents share a model for jointly interpreting symbols and inferring actions.
The model is specified as follows:
\begin{align}
  &\text{prior:} & P(s^*_t \mid s^*_{t-1}, a^*_{t-1}) &= \mathrm{Cat}(\mathbf{B}^*_{a_{t-1}}), \label{eq:prior} \\
  &\text{initial prior:} & P(s^*_1) &= \mathrm{Cat}(\mathbf{D}^*), \label{eq:init_prior} \\
  &\text{posterior:} & Q(s^*_t) &= \mathrm{Cat}(\phi^*_t), \label{eq:posterior} \\
  &\text{likelihood:} & P(o^*_t \mid s^*_t) &= \mathrm{Cat}(\mathbf{A}^*), \label{eq:likelihood} \\
  &\text{preference:} & \tilde{P}(o^*_t) &= \mathrm{Cat}(\mathbf{C}^*), \label{eq:preference} \\
  &\text{symbol inference:} & P(w^*_t) &= \mathrm{Cat}(\xi^*_t), \label{eq:symbol_inf} \\
  &\text{symbol interpretation:} & \pi(a^*_t \mid w^*_t) &= \mathrm{Cat}(\mathbf{E}), \label{eq:symbol_interp}
\end{align}
where $* \in \{A, B\}$ indexes the agent.
Let $N_o$, $N_s$, $N_a$, and $N_w$ denote the numbers of possible observations, states, actions, and symbols, respectively.
The observation $o^*_t$ and the state $s^*_t$ of agent~$*$ are each represented as one-hot column vectors of dimensions $N_o$ and $N_s$.
The parameters $\mathbf{A}^* \in \mathbb{R}^{N_o \times N_s}$, $\mathbf{B}^*_a \in \mathbb{R}^{N_s \times N_s}$, $\mathbf{D}^* \in \mathbb{R}^{N_s}$, $\mathbf{C}^* \in \mathbb{R}^{N_o}$, and $\mathbf{E} \in \mathbb{R}^{N_a \times N_w}$ correspond to the categorical distributions over the likelihood, state transition, initial state, prior preference, and symbol interpretation, respectively.
A separate transition matrix $\mathbf{B}^*_a$ exists for each action; we write $\mathbf{B}^*$ to refer to the collection of all such matrices.
In this model, $\mathbf{B}^*$ and $\mathbf{D}^*$ are given accurate values in advance, while the remaining parameters $\mathbf{A}^*$, $\mathbf{C}^*$, and $\mathbf{E}$ are learned through online updates.

In this model, the symbols inferred from bodily states and prior preferences are regarded as socially shared emotion concepts.
Social reality refers to the state in which emotion concepts have emerged in a form agreed upon within the community.
Concretely, this model treats as social reality the condition in which the prior preferences $\mathbf{C}^*$ of the agents have become similar and the symbol interpretation parameter $\mathbf{E}$, which encodes the actions for controlling bodily states so as to satisfy those preferences, has been formed.

\subsection{Variational Bayesian Inference}
\label{sec:model:vb}

Each agent infers its state by finding the approximate posterior parameter $\phi^*_t$ that minimizes the variational free energy $F^*$ under active inference.
Following~\cite{Da-Costa2020-po}, the parameter $\phi^*_t$ that minimizes the variational free energy of agent~$*$ is obtained as
\begin{equation}
  \phi^*_t \approx \sigma\!\bigl(\log(\mathbf{A}^* \cdot o^*_t) + \log(\mathbf{B}^*_{a_{t-1}} \phi^*_{t-1})\bigr),
  \label{eq:vfe}
\end{equation}
where $\sigma$ denotes the softmax function.

\subsection{Symbol Sampling Based on Active Inference and Collective Predictive Coding}
\label{sec:model:symbol}

Each agent infers symbols through active inference.
However, because neither agent can determine on its own an action that satisfies both agents' prior preferences, the symbol actually used for action selection is sampled through the Metropolis--Hastings Naming Game (MHNG).

\subsubsection{Symbol Inference via Active Inference}
\label{sec:model:symbol:ai}

Active inference selects actions that minimize the expected free energy.
In this model, where inferred symbols are interpreted to determine actions, symbols must be inferred so that the resulting actions are likely to minimize the expected free energy.
Accordingly, the model infers the symbol whose expected value of expected free energy, computed over the actions derived from each symbol, is minimized.
The expected free energy for action $a^*_t$ of agent~$*$ is given by
\begin{align}
  G^*(a^*_t)
  &\approx \mathbb{E}_{Q(s^*_{t+1}, o^*_{t+1} \mid a^*_t)}
    \bigl[\log Q(o^*_{t+1} \mid a^*_t)
    - \log Q(o^*_{t+1} \mid s^*_{t+1}, a^*_t)
    - \log \tilde{P}(o^*_{t+1})\bigr] \notag \\
  &= (\mathbf{B}^* \phi^*_t) \cdot H[\mathbf{A}^*]
    + \bigl\{\log(\mathbf{A}^* \mathbf{B}^* \phi^*_t - \mathbf{C}^*)\bigr\} \cdot \mathbf{A}^* \mathbf{B}^* \phi^*_t.
  \label{eq:efe}
\end{align}
Let $g^*_t$ denote the column vector obtained by stacking the expected free energies computed for each action.
The expected value of the expected free energy for each symbol is then
\begin{equation}
  G^*_{\mathrm{symbol}} = \mathbf{E} \cdot g^*_t.
  \label{eq:efe_symbol}
\end{equation}
The parameter of the symbol inference distribution is therefore computed as
\begin{equation}
  \xi^*_t = \sigma(-G^*_{\mathrm{symbol}}).
  \label{eq:xi}
\end{equation}

\subsubsection{Symbol Sampling via the Metropolis--Hastings Naming Game}
\label{sec:model:symbol:mhng}

At each time step, both agents infer symbols, and the MHNG determines which agent's symbol is used for action selection.
The goal of symbol emergence in this model is to infer symbols that minimize the expected value of expected free energy for both agents simultaneously.
However, because each agent is confined to its own sensorimotor system and operates as an independent individual, neither can access the other's internal information.
The MHNG therefore provides a mechanism whereby an agent either accepts the symbol proposed by the other agent or rejects it in favor of its own inferred symbol, and the selected symbol is then used to infer the action.

The MHNG uses the Metropolis--Hastings algorithm to draw samples from the target distribution $P(w_t; \bar{\xi}_t)$.
Using the Product-of-Experts approximation, the target distribution is approximated as
\begin{equation}
  P(w_t; \bar{\xi}_t) \approx \frac{1}{Z}\,P(w_t; \xi^A_t)\,P(w_t; \xi^B_t),
  \label{eq:mhng_target}
\end{equation}
where $Z$ is the normalizing constant.
The agent that selects a symbol and acts is called the Listener, and the agent that proposes a symbol is called the Speaker; their subscripts are denoted $\mathrm{Li}$ and $\mathrm{Sp}$, respectively.
The proposal distribution is $P(w^{\mathrm{Sp}}_t; \xi^{\mathrm{Sp}}_t)$.
The acceptance rate of $w^{\mathrm{Sp}}_t$ in the Metropolis--Hastings algorithm is then
\begin{equation}
  r = \frac{Q(w^{\mathrm{Li}}_t)\,P(w^{\mathrm{Sp}}_t)}{Q(w^{\mathrm{Sp}}_t)\,P(w^{\mathrm{Li}}_t)}
    = \frac{P(w^{\mathrm{Sp}}_t;\, \xi^{\mathrm{Li}}_t)}{P(w^{\mathrm{Li}}_t;\, \xi^{\mathrm{Li}}_t)}.
  \label{eq:acceptance}
\end{equation}
This expression shows that the decision to accept or reject the Speaker's proposed symbol can be computed solely from the Listener's information.

\subsection{Parameter Updates}
\label{sec:model:update}

\subsubsection{Update of the Likelihood Parameter $\mathbf{A}^*$}
\label{sec:model:update:A}

The likelihood parameter $\mathbf{A}^*$ of each agent is updated by sequentially updating online the parameter of the Dirichlet distribution from which $\mathbf{A}^*$ is drawn.
Let $\mathbf{a}^*$ denote the Dirichlet parameter.
The Dirichlet parameter is updated so as to minimize the variational free energy with respect to the generative model parameters.
Following~\cite{Da-Costa2020-po}, the Bayesian update of the Dirichlet parameter is
\begin{equation}
  \mathbf{a}^* \leftarrow \mathbf{a}^* + o^*_t \otimes \phi^*_t.
  \label{eq:update_A}
\end{equation}

\subsubsection{Update of the Prior Preference Parameter $\mathbf{C}^*$}
\label{sec:model:update:C}

This model realizes allostasis of bodily states that reflects social constraints mediated by symbols, by mutually updating the prior preferences when the symbol inferred by one agent is rejected as the other agent controls its bodily state.
When the two agents hold different prior preferences, their expected free energy landscapes differ.
Different expected free energy landscapes lead to different preferred actions for each agent, causing the agents to infer different symbols corresponding to those actions.
Consequently, it becomes difficult to take actions that transition to states satisfying both agents' preferences simultaneously, and the rejection rate of symbols in the MHNG increases.

To address this, under the condition that the mean entropy of the likelihood distribution $\mathbb{E}[H[\mathbf{A}^*]]$ contained in the first term of the expected free energy is sufficiently small---specifically, below a threshold $H_{\mathrm{thres}}$---the Listener updates its prior preference when the MHNG rejects its proposed symbol.
The update moves the preference toward the observation distribution predicted when using the Listener's accepted symbol (i.e., the symbol inferred by the other agent) and away from the distribution predicted when using the Speaker's own proposed symbol.
The observation distribution given symbol $w^*_t$ is computed by marginalizing over actions as
\begin{equation}
\begin{aligned}
P(o^\ast_{t+1} \mid w^\ast_t)
&=
\sum_{a^\ast}
\mathbf{E}_{a^\ast,w^\ast_t}
\mathbf{A}^\ast \mathbf{B}^\ast_{a^\ast}\phi^\ast_t
\\
&=
\left[
\mathbf{A}^\ast \mathbf{B}^\ast_{a_1}\phi^\ast_t,\,
\ldots,\,
\mathbf{A}^\ast \mathbf{B}^\ast_{a_{N_a}}\phi^\ast_t
\right]
\mathbf{E}_{:,w^\ast_t}
\\
&=
\mathbf{A}^\ast \mathbf{B}^\ast\phi^\ast_t \mathbf{E}_{:,w^\ast_t}.
\label{eq:predicted_observation_given_symbol}
\end{aligned}
\end{equation}
Let $\tilde{\mathbf{C}}^*$ denote the preference score representing the evaluation of each observation, initialized as $\tilde{\mathbf{C}}^* = \log \mathbf{C}^* - \bar{\mathbf{C}}^*$, where $\bar{\mathbf{C}}^*$ is the mean of $\mathbf{C}^*$.
The preference score is updated as follows, and the prior preference parameter is obtained by applying the softmax function:
\begin{align}
  \tilde{\mathbf{C}}^{\mathrm{Sp}} &\leftarrow \tilde{\mathbf{C}}^{\mathrm{Sp}}
    + \eta_C \bigl(P(o^{\mathrm{Sp}}_{t+1} \mid w^{\mathrm{Li}}_t) - P(o^{\mathrm{Sp}}_{t+1} \mid w^{\mathrm{Sp}}_t)\bigr) \notag \\
  &= \tilde{\mathbf{C}}^{\mathrm{Sp}}
    + \eta_C \bigl((\mathbf{A}^{\mathrm{Sp}} \mathbf{B}^{\mathrm{Sp}} \phi^{\mathrm{Sp}}_t \mathbf{E})_{w^{\mathrm{Li}}_t}
    - (\mathbf{A}^{\mathrm{Sp}} \mathbf{B}^{\mathrm{Sp}} \phi^{\mathrm{Sp}}_t \mathbf{E})_{w^{\mathrm{Sp}}_t}\bigr), \label{eq:update_C} \\
  \mathbf{C}^{\mathrm{Sp}} &= \sigma(\tilde{\mathbf{C}}^{\mathrm{Sp}}), \notag
\end{align}
where $\eta_C$ is the learning rate for the prior preference.

\subsubsection{Update of the Symbol Interpretation Parameter $\mathbf{E}$}
\label{sec:model:update:E}

The symbol interpretation parameter $\mathbf{E}$, which maps symbols to actions, is updated so that the action minimizing the expected free energy is inferred from the symbol sampled in the MHNG.
Suppose the symbol $w^*_t$ is sampled; then the update is
\begin{equation}
  \mathbf{E}_{w^*_t} \leftarrow (1 - \eta_E)\,\mathbf{E}_{w^*_t} + \eta_E\,\sigma(-g^*_t / \tau),
  \label{eq:update_E}
\end{equation}
where $\eta_E$ is the learning rate for symbol interpretation and $\tau$ is a temperature parameter.

\subsection{Online Inference and Update}
\label{sec:model:online}

At every time step, the model infers states and symbols from the observed interoceptive signal, performs communication via the MHNG, and simultaneously updates the model parameters online.
The prior preference parameter $\mathbf{C}^*$ and the symbol interpretation parameter $\mathbf{E}$ are not updated simultaneously; instead, they are updated alternately, switching every $T_{\mathrm{phase}}$ steps.

At each step, the agents take turns as Listener and Speaker.
The Listener infers an action using the symbol sampled through the MHNG, and the Speaker updates the prior preference parameter $\mathbf{C}^{\mathrm{Sp}}$ when the update condition is satisfied.

\section{Experiments}
\label{sec:experiments}

\subsection{Experimental Setup}
\label{sec:experiments:setup}

The experiment examines whether social reality---a state in which the prior preferences over interoceptive sensations are aligned between the two agents and symbols representing actions for bodily state control have been formed---emerges through the proposed model.
Bodily states are represented by two modalities: energy and body temperature.
Each agent receives discrete multimodal interoceptive signals composed of energy and body temperature values.
Energy is expressed on a six-level scale from 0 to 5, where higher values indicate greater accumulation.
Body temperature is similarly expressed on a six-level scale from 0 to 5, where higher values indicate higher temperature.
Energy and body temperature are assumed to change independently.
The interoceptive signal therefore takes $6 \times 6 = 36$ possible values and is represented as a 36-dimensional one-hot vector.

Each agent can take one of five actions: Cool, Warm, Eat, Play, and Sleep.
These actions affect the interoceptive sensations as follows.
Cool decreases energy by 1 and body temperature by 1.
Warm decreases energy by 1 and increases body temperature by 1.
Eat increases energy by 2, and with 20\% probability changes body temperature by $-1$ if the current temperature is 2 or below or by $+1$ if it is 3 or above.
Play decreases energy by 1, and with 20\% probability changes body temperature by $-1$ if the current temperature is 2 or below or by $+1$ if it is 3 or above.
Sleep has no effect on energy, and with 20\% probability changes body temperature by $-1$ if the current temperature is 2 or below or by $+1$ if it is 3 or above.

The dimensions $N_s$ and $N_w$ were set to 36 and 15, respectively.
The initial value of $\mathbf{A}^*$ was set to a uniform distribution.
The initial prior preference of each agent was configured so that both agents prefer a moderate energy level, while Agent~A prefers high body temperature and Agent~B prefers low body temperature (see the leftmost panel of Fig.~\ref{fig:preference}).
The symbol interpretation parameter $\mathbf{E}$ was initialized so that all actions have equal expected probability (see the leftmost panel of Fig.~\ref{fig:interpretation}).

The experimental scenario involves two agents that start with different prior preferences and no prior knowledge about symbols, and that control their bodily states through active inference while communicating.
The experiment is designed to test the following two hypotheses.
First, updating prior preferences in response to the other agent's symbol when one's own symbol proposal is rejected leads to mutual adaptation, resulting in convergence to similar prior preferences.
Second, symbols that represent the actions needed to satisfy the prior preferences, together with a symbol interpretation that maps those symbols to actions, emerge through the interaction.

\subsection{Results}
\label{sec:experiments:results}

\subsubsection{Changes in Prior Preferences}
\label{sec:experiments:results:C}

To examine whether the prior preferences of the two agents converge as the model infers symbols and actions and updates its parameters, we analyzed the change in the distance between the two agents' prior preference distributions.
The Jensen--Shannon (JS) divergence was used as the distance measure.

\begin{figure}[t]
    \centering
    \includegraphics[width=0.6\linewidth]{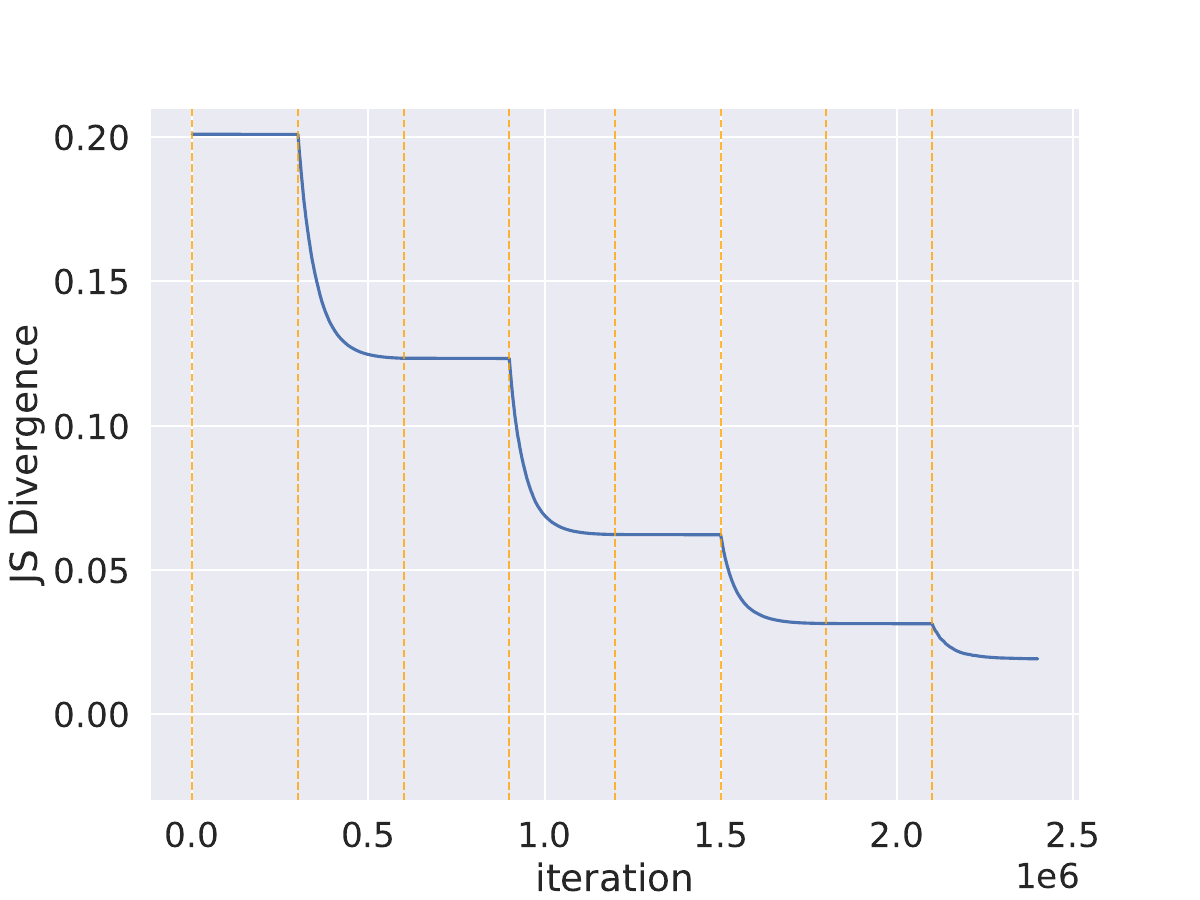}
    \caption{Trajectory of the Jensen--Shannon divergence between the prior preference distributions of the two agents over iterations.
    The decreasing divergence indicates that the agents' prior preferences gradually converge through symbolic communication.}
    \label{fig:jsd}
\end{figure}

Figure~\ref{fig:jsd} shows the trajectory of the JS divergence between the prior preferences.
The JS divergence decreases as the iteration count increases, confirming that the two agents' prior preferences gradually become more similar.
In this model, when a proposed symbol is rejected, the proposing agent shifts its prior preference toward the observation probability predicted by the symbol accepted by the other agent.
This mechanism enables each agent to form a prior preference adapted to the other agent in a top-down manner through communication, without directly accessing the other agent's preference information.

\begin{figure}[t]
    \centering
    \includegraphics[width=0.9\linewidth]{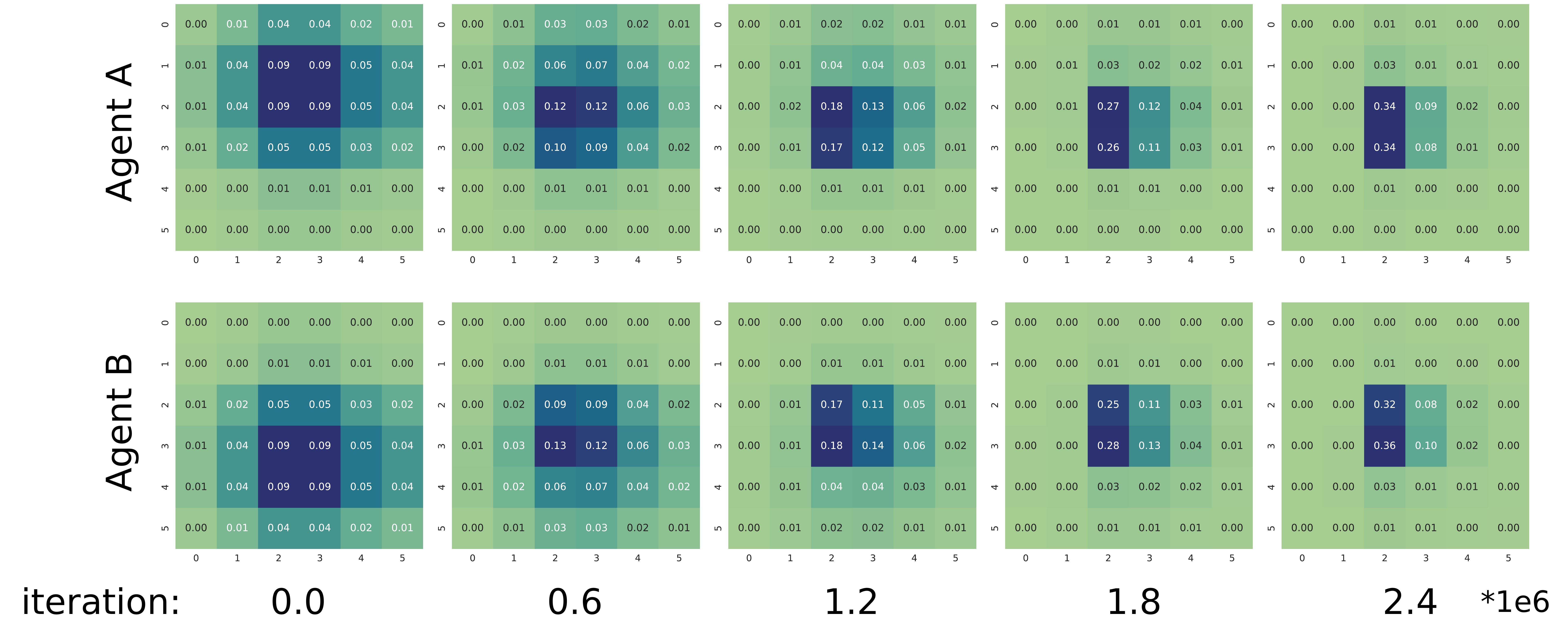}
    \caption{Change in the prior preference parameter $\mathbf{C}^*$ of each agent over iterations.
    Each heatmap shows the preference distribution over the $6 \times 6$ interoceptive state space (energy $\times$ body temperature).
    Starting from different initial preferences (Agent~A prefers high temperature; Agent~B prefers low temperature), both agents' preferences converge toward an intermediate configuration.}
    \label{fig:preference}
\end{figure}

Figure~\ref{fig:preference} shows the change in the prior preference parameter $\mathbf{C}^*$ of each agent.
Starting from different initial preferences, the two agents' preferences converge over iterations so that the peak is located at an intermediate position between the two initial preferences.
This result indicates that the proposed method causes each agent to shift its prior preference so that its inferred symbols are more likely to be accepted by the other agent, and consequently a prior preference that can simultaneously satisfy both agents is formed.

\subsubsection{Formation of Symbol Interpretation}
\label{sec:experiments:results:E}

\begin{figure}[t]
    \centering
    \includegraphics[width=0.85\linewidth]{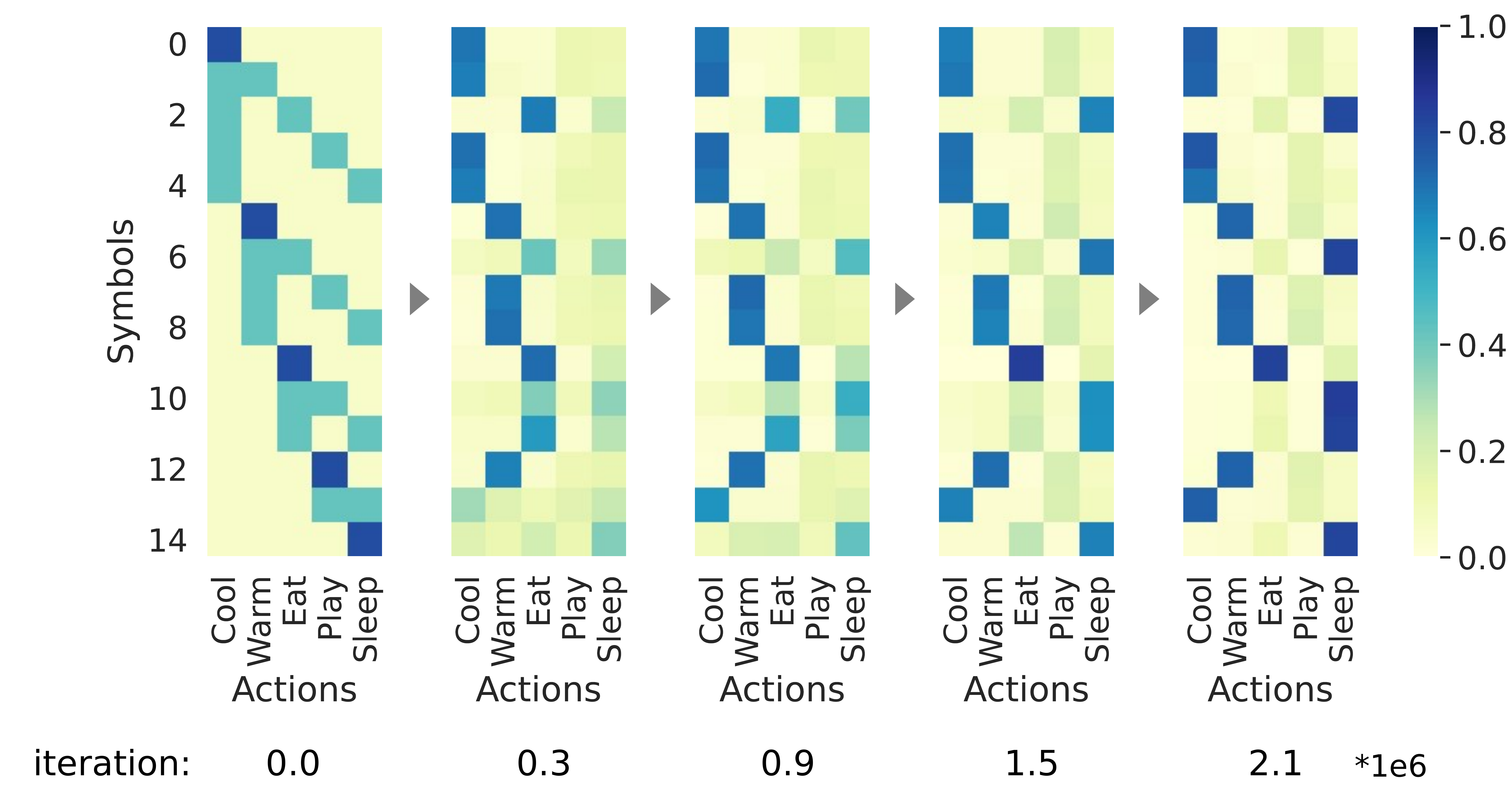}
    \caption{Change in the symbol interpretation parameter $\mathbf{E}$ over iterations.
    Rows correspond to the 15 symbols and columns correspond to the five actions (Cool, Warm, Eat, Play, Sleep).
    Darker cells indicate higher probability.
    In the early phase, symbols encoding temperature-regulation and energy-replenishing actions predominate; in the later phase, some symbols shift from Eat to Sleep as the agents' preferences become aligned.}
    \label{fig:interpretation}
\end{figure}

To examine whether a symbol interpretation for inferring actions from symbols has emerged, we analyzed the change in the symbol interpretation parameter $\mathbf{E}$.
Figure~\ref{fig:interpretation} shows the transition of $\mathbf{E}$.
From the initial to the middle phase, symbols representing Cool, Warm, and Eat increase in number.
This is likely because the discrepancy in the two agents' temperature-related prior preferences during the early phase leads to frequent temperature-regulation actions, which in turn necessitate compensating for the resulting energy expenditure.
From the middle to the final phase, the agents' prior preferences become aligned and the need for energy expenditure decreases; as a result, symbols that previously represented Eat shift to represent Sleep.
These observations confirm that the symbol interpretation is modified with the objective of selecting actions that minimize the expected free energy, thereby changing the meaning carried by each symbol.

\section{Discussion}
\label{sec:discussion}

The experimental results support the two hypotheses set out in Section~\ref{sec:experiments:setup}.
The decrease in JS divergence between the two agents' prior preferences over iterations demonstrates that the proposed mutual update mechanism enables agents to align their bodily control goals through symbolic communication alone, without direct access to each other's internal states.
The convergence of the prior preference parameters to an intermediate position between the two initial configurations indicates that the alignment is not one-sided but rather a compromise that accommodates both agents' original preferences.

The analysis of the symbol interpretation parameter $\mathbf{E}$ reveals that the meanings of symbols are not static but evolve in response to changes in the agents' internal states and preferences.
During the early phase, when the agents' temperature preferences diverge, symbols encoding temperature-regulation actions (Cool, Warm) and energy-replenishing actions (Eat) are predominantly formed.
As the preferences become aligned in the later phase and energy demands decrease, some symbols shift their meaning from Eat to Sleep.
This dynamic reassignment of symbol meaning is consistent with the view in the theory of constructed emotion that emotion concepts are not fixed but are continuously negotiated through social interaction.

The proposed model has several limitations.
The present study is restricted to a discrete environment and a discrete generative model with a simple configuration.
In future work, we plan to extend the model to continuous environments and more complex settings so as to examine the robustness and scalability of the social reality emergence mechanism.

\section{Conclusion}
\label{sec:conclusion}

This paper proposed a model for the emergence of social reality of emotion, in which two agents equipped with POMDP-based generative models of interoceptive sensations infer symbols and actions through active inference and MHNG to mutually control their bodily states.
By updating one's own prior preference to accommodate the other agent on the basis of symbol rejection in the MHNG, the model enables each agent to acquire a prior preference adapted to the other agent without directly referencing the other's preference information.
In addition, by modifying the symbol interpretation so that actions minimizing the expected free energy are more readily inferred, the model demonstrates that the meanings of symbols shared between the agents can change dynamically.

\begin{ack}
This work was supported by JSPS KAKENHI Grant Number JP23H04834.
\end{ack}

\bibliographystyle{plain}
\bibliography{reference}

\end{document}

%% file: math_commands.tex
\usepackage{amsmath,amsfonts,bm}

\def\eqref#1{equation~\ref{#1}}

\def\1{\bm{1}}

\DeclareMathAlphabet{\mathsfit}{\encodingdefault}{\sfdefault}{m}{sl}
\SetMathAlphabet{\mathsfit}{bold}{\encodingdefault}{\sfdefault}{bx}{n}

